\documentclass[pre,twocolumn,superscriptaddress]{revtex4-1}
\usepackage[dvipdfmx]{graphicx}
\usepackage{type1cm}
\usepackage[usenames,svgnames]{xcolor}
\usepackage{color}
\usepackage{url}
\usepackage{natbib}
\usepackage[dvipdfmx]{hyperref}
\hypersetup{
     colorlinks=true,       		
     linkcolor=Navy,          	
     citecolor=Navy,            
     filecolor=Navy,      		
     urlcolor=Navy,           	
    runcolor=cyan,
 }

\usepackage{amsmath,amssymb,bm,amsthm}

\def\beq{\begin{equation}}
\def\eeq{\end{equation}}
\def\nbeq{\begin{equation*}}
\def\neeq{\end{equation*}}

\def\<{\langle}
\def\>{\rangle}

\newcommand{\sectionprl}[1]{{\par\it #1.---}}

\begin{document}
\title{Macrostate equivalence of two general ensembles and specific relative entropies}

\author{Takashi Mori}
\affiliation{
Department of Physics, Graduate School of Science,
University of Tokyo, Bunkyo-ku, Tokyo 113-0033, Japan
}

\begin{abstract}
The two criteria of ensemble equivalence, i.e. the macrostate equivalence and the measure equivalence, are investigated for a general pair of states. 
The macrostate equivalence implies the two ensembles are indistinguishable by the measurement of macroscopic quantities obeying the large-deviation principle, and the measure equivalence means that the specific relative entropy of these two states vanishes in the thermodynamic limit.
It is shown that the measure equivalence implies the macrostate equivalence for a general pair of states by deriving an inequality connecting the large-deviation rate functions to the specific relative Renyi entropies.
The result is applicable to both quantum and classical systems.
As applications, a sufficient condition for thermalization, the timescale of quantum dynamics of macrovariables, and the second law with strict irreversibility in a quantum quench are discussed.
\end{abstract}
\maketitle

Ensemble equivalence in the thermodynamic limit is a fundamental result of equilibrium statistical mechanics~\cite{Ruelle_text}.
The concavity of the microcanonical entropy implies the \textit{thermodynamic equivalence} among the microcanonical, the canonical, and the grandcanonical ensembles;
the thermodynamic functions can be transformed with each other via the Legendre transformation.
The ensemble equivalence has been also studied at the deeper levels: the \textit{macrostate equivalence}~\cite{Ellis2000,Touchette2004} and the \textit{measure equivalence}~\cite{Lewis1994}.
Roughly speaking, macrostate equivalence means that the two ensembles possess the identical set of equilibrium values of macrovariables, and the measure equivalence means that the specific relative entropy between the two ensembles vanishes in the thermodynamic limit.
When we consider the conventionally-used thermal ensembles, i.e. microcanonical, canonical, and grandcanonical ensembles, it is rigorously shown that these three levels of ensemble equivalence coincide with each other~\cite{Touchette2015}.

From a broader perspective beyond the equilibrium statistical mechanics, we encounter many different kinds of statistical ensembles.
In the study of relaxation in isolated quantum systems, it has been recognized that the typicality of the equilibrium state is essential in the foundation of statistical mechanics~\cite{Neumann1929,Tasaki1998,Goldstein2006,Reimann2007,Goldstein2010, Tasaki2016_typicality}.
Typicality of the equilibrium state indicates that there are a lot of statistical ensembles which are not conventionally used but are macroscopically indistinguishable from the microcanonical ensemble.
The stationary state starting from a nonequilibrium initial state is described by the diagonal ensemble~\cite{Rigol2008}.
It is therefore of great importance in the studies of thermalization to understand under what condition the diagonal ensemble is equivalent to the microcanonical ensemble.
In integrable systems, it is conjectured that the steady state is described by the generalized Gibbs ensemble, in which an extensive number of local conserved quantities are taken into account~\cite{Rigol2007,Cassidy2011,Essler2015,Ilievski2015,Yuzbashyan2016}.
Needless to say, in driven diffusive systems, nonequilibrium steady states are not described by thermal ensembles~\cite{Katz1984,Derrida2002,Derrida2007,Cohen-Mukamel2012}.

Even in these cases, the notions of the macrostate equivalence and the measure equivalence do not lose their meaning.
In particular, the macrostate equivalence has a clear physical meaning, that is, it requires that two ensembles are indistinguishable as long as we consider macroscopic observables.
However, it is difficult to confirm that the macrostate equivalence holds for a concrete pair of states, because we must check that the typical values of any macrovariable are same in the two ensembles.
Therefore, it is a challenge to explore the condition of the macrostate equivalence between the two general states, not necessarily conventional thermal ensembles.
With this regard, the relative entropy is sometimes easier to calculate, but a physical meaning of the measure equivalence is not so clear as it is.

In this paper, it is shown that we can extract useful information on the large-deviation rate functions in two general states from the relative Renyi entropies.
It is done by deriving the inequality relating the specific relative Renyi entropy to the large-deviation rate function of a macrovariable.
This inequality shows that the vanishing specific relative entropy in the thermodynamic limit, i.e. the measure equivalence, implies the macrostate equivalence.
After presenting the main result, we discuss an immediate application to the large-deviation probabilities in the microcanonical and the canonical ensembles.
Then, we discuss the problem of thermalization and give a sufficient condition for thermalization, and the result is also applied to an estimation of timescale of quantum dynamics of macrovariables.
Finally, the second law of thermodynamics with strict irreversibility in a quantum quench is derived.

\sectionprl{Setting}
We consider sequences of density matrices $(\rho_N)_{N\in\mathbb{N}}$ and $(\tau_N)_{N\in\mathbb{N}}$ in a quantum system with $N$ particles or spins.
We can also consider the case of large but finite fixed $N$, but the situation becomes complicated, so in this paper, for simplicity, we focus on the case in which there are proper thermodynamic limits of $\rho_N$ and $\tau_N$.
The case of large but finite fixed $N$ will be discussed elsewhere.
The (quantum) relative entropy~\cite{Nielsen-Chuang_text} of $\tau_N$ with respect to $\rho_N$ is given by
\beq
S(\tau_N\|\rho_N)={\rm Tr}\tau_N(\ln\tau_N-\ln\rho_N).
\label{eq:relative}
\eeq
The specific relative entropy in the thermodynamic limit is defined as
$s(\tau\|\rho)=\lim_{N\rightarrow\infty}S(\tau_N\|\rho_N)/N$.
When $s(\tau\|\rho)=0$ or $s(\rho\|\tau)=0$, the states $\tau$ and $\rho$ are said to be measure equivalent.
We also introduce the (quantum) relative Renyi entropy~\cite{Renyi1961} as
\beq
S_{\alpha}(\tau_N\|\rho_N)=\frac{1}{\alpha-1}\ln{\rm Tr}\tau_N^{\alpha}\rho_N^{1-\alpha}.
\label{eq:Renyi}
\eeq
The relative entropy is obtained by $S(\tau_N\|\rho_N)=\lim_{\alpha\rightarrow 1}S_{\alpha}(\tau_N\|\rho_N)$.
The specific relative Renyi entropy is defined as $s_{\alpha}(\tau\|\rho)=\lim_{N\rightarrow\infty}S_{\alpha}(\tau_N\|\rho_N)/N$.
It is known that $S_{\alpha}(\tau_N\|\rho_N)$ is non-negative and non-decreasing with $\alpha$.
In this paper, we consider quantum systems, but we can also discuss classical systems by considering the diagonal density matrices $\rho_N$ and $\tau_N$.

We consider a ``macrovariable'' $X$, which is an Hermitian operator.
Eigenvectors of the operator $X$ are defined as $X|X_i\>=X_i|X_i\>$ with eigenvalues $X_1\leq X_2\leq\dots$.
We define the probability of finding a measurement outcome of $X$ in $[x,x')$ in the state $\sigma_N=\rho_N$ or $\tau_N$ as
\beq
P_{\sigma_N}(X\in[x,x')):={\rm Tr}\sigma_N\mathsf{P}_{X\in[x,x')},
\eeq
where $\mathsf{P}_{X\in[x,x')}:=\sum_{i:X_i\in[x,x')}|X_i\>\<X_i|$ is the projection onto the subspace spanned by $\{|X_i\>:X_i\in[x,x')\}$.
We define the rate function~\footnote{We call Eq.~(\ref{eq:rate}) a ``rate function'' although the rate function is defined by using $\lim_{N\rightarrow\infty}$ instead of $\limsup_{N\rightarrow\infty}$ in Eq.~(\ref{eq:rate}) in the theory of large deviations.}
of $X$ in the state $\sigma$ as
\beq
I_{\sigma}(x):=-\limsup_{N\rightarrow\infty}\frac{\ln P_{\sigma_N}(X\in[x,x+dx))}{N},
\label{eq:rate}
\eeq
where $dx$ is infinitesimal but it is implicitly assumed that the limit of $dx\rightarrow +0$ is always taken after the limit of $N\rightarrow\infty$ in this paper.
A positive rate function $I_{\sigma}(x)>0$ implies the exponentially small probability of finding the measurement outcome $x$ in a large system.
We define the set of ``typical values'' as
\beq
\mathcal{E}_{\sigma}:=\{x\in\mathbb{R}:I_{\sigma}(x)=0\}.
\eeq
The macrostate equivalence between the states $\rho$ and $\tau$ means $\mathcal{E}_{\rho}=\mathcal{E}_{\tau}$.
The macrostate partial equivalence means $\mathcal{E}_{\rho}\supset\mathcal{E}_{\tau}$ or $\mathcal{E}_{\tau}\supset\mathcal{E}_{\rho}$.
Otherwise, the two states $\rho$ and $\tau$ are macrostate nonequivalent.

The rate function defined above exists for any observable, which is not necessarily a ``macrovariable''.
However, for an observable with large fluctuations, $I_{\sigma}(x)=0$ for a very wide range of $x$ and the meaning of $\mathcal{E}_{\sigma}$ as the set of typical values is lost.
We shall say that $X$ is a macrovariable if $\mathcal{E}_{\sigma}$ is a discrete set and the number of its element is finite (therefore, we do not consider the case of phase coexistence).
The inequalities we will show below are applicable to any observable $X$, but they contain particularly useful information when they are applied to macrovariables.

In equilibrium states, we believe that thermodynamic quantities satisfy the large deviation principle except for the critical point.
For quantum lattice systems in equilibrium states~\cite{Netocny2004,Lenci2005,Ogata2010}, it is partially proved that there exists the rate function with a single typical value for quantities defined as the spatial average of a local operator, $X=(1/N)\sum_{i=1}^NO_i$, where $i$ denotes each site and $O_i$ is the translation of $O_0$, which is a local operator acting onto a finite number of sites (0 stands for the origin).
We call $X$ expressed as $X=(1/N)\sum_{i=1}^NO_i$ a local intensive variable~\cite{Biroli2010}.
Also in nonequilibrium states, it is expected that the large deviation principle holds for ``macroscopic quantities'' in many cases~\cite{Derrida2002,Derrida2007,Cohen-Mukamel2012}.
For simplicity, we consider a single macrovariable, but generalization to multiple macrovariables are straightforward.

\sectionprl{Inequalities for the rate functions}
We discuss the relation between the macrostate equivalence and the measure equivalence based on the following inequality:
\beq
I_{\tau}(x)\geq \frac{\alpha-1}{\alpha}\left[I_{\rho}(x)-s_{\alpha}(\tau\|\rho)\right]
\label{eq:main}
\eeq
for any $\alpha>1$.
This inequality implies that
\beq
\mathcal{E}_{\tau}\subseteq \{x: I_{\rho}(x)\leq s_{\alpha}(\tau\|\rho)\}.
\label{eq:macro_alpha}
\eeq
By taking the limit of $\alpha\rightarrow 1$ from above, we have
\beq
\mathcal{E}_{\tau}\subseteq \{x:I_{\rho}(x)\leq s(\tau\|\rho)\}.
\label{eq:macro}
\eeq

When $\rho$ and $\tau$ are measure equivalent, $s(\tau\|\rho)=0$, Eq.~(\ref{eq:macro}) shows $\mathcal{E}_{\tau}\subseteq\mathcal{E}_{\rho}$, that is, the states $\rho$ and $\tau$ are at least partially equivalent in the macrostate level.
If the typical value in the state $\rho$ is unique, $\mathcal{E}=\{\<X\>_{\rho}\}$, where $\<\cdot\>_{\sigma}:={\rm Tr}\sigma(\cdot)$, or if both of $s(\tau\|\rho)$ and $s(\rho\|\tau)$ are zero, then we have the macrostate equivalence $\mathcal{E}_{\rho}=\mathcal{E}_{\tau}$.
In this sense, we could show that the measure equivalence implies the macrostate equivalence between general states $\rho$ and $\tau$.

The result~(\ref{eq:macro}) is useful even if $s(\tau\|\rho)$ does not vanish.
Equation~(\ref{eq:macro}) tells us that the possible outcome of measuring $X$ must be in the region with $I_{\rho}(x)\leq s(\tau\|\rho)$.
If we regard $\tau$ as an approximation of $\rho$, Eq.~(\ref{eq:macro}) gives an estimate of error due to the approximation.

\begin{proof}[Proof of Eq.~(\ref{eq:main})]
For notational simplicity, here we write $\mathsf{P}_{X\in[x,x+dx)}=\mathsf{P}$, $\tau_N=\tau$, and $\rho_N=\rho$.
We have
\begin{align}
&P_{\tau}(X\in[x,x+dx))={\rm Tr}\tau\mathsf{P}
\nonumber \\
&={\rm Tr}\rho^{-\frac{\alpha-1}{2\alpha}}\tau\rho^{-\frac{\alpha-1}{2\alpha}}\cdot\rho^{\frac{\alpha-1}{2\alpha}}\mathsf{P}\rho^{\frac{\alpha-1}{2\alpha}}
\end{align}
with $\alpha>1$.
By applying the H\"older's inequality,
\begin{align}
P_{\tau}(X\in[x,x+dx))
\leq\left[{\rm Tr}\left(\rho^{-\frac{\alpha-1}{2\alpha}}\tau\rho^{-\frac{\alpha-1}{2\alpha}}\right)^{\alpha}\right]^{\frac{1}{\alpha}}
\nonumber \\
\times \left[{\rm Tr}\left(\rho^{\frac{\alpha-1}{2\alpha}}\mathsf{P}\rho^{\frac{\alpha-1}{2\alpha}}\right)^{\frac{\alpha}{\alpha-1}}\right]^{\frac{\alpha-1}{\alpha}}.
\end{align}
By applying the Araki-Lieb-Thirring inequality~\cite{Araki1990,Lieb-Thirring_inequalities}, we have
\beq
\left[{\rm Tr}\left(\rho^{-\frac{\alpha-1}{2\alpha}}\tau\rho^{-\frac{\alpha-1}{2\alpha}}\right)^{\alpha}\right]^{\frac{1}{\alpha}}\leq \left({\rm Tr}\tau^{\alpha}\rho^{1-\alpha}\right)^{\frac{1}{\alpha}}
=e^{\frac{\alpha-1}{\alpha}S_{\alpha}(\tau\|\rho)}
\label{eq:proof1}
\eeq
and
\beq
\left[{\rm Tr}\left(\rho^{\frac{\alpha-1}{2\alpha}}\mathsf{P}\rho^{\frac{\alpha-1}{2\alpha}}\right)^{\frac{\alpha}{\alpha-1}}\right]^{\frac{\alpha-1}{\alpha}}
\leq \left({\rm Tr}\rho\mathsf{P}\right)^{\frac{\alpha-1}{\alpha}}
\sim e^{-\frac{\alpha-1}{\alpha}NI_{\rho}(x)}.
\eeq
Thus we have
$I_{\tau}(x)=-\limsup_{N\rightarrow\infty}\ln P_{\tau}(X\in[x,x+dx))/N
\geq[(\alpha-1)/\alpha]\left[I_{\rho}(x)-s_{\alpha}(\tau\|\rho)\right]$.
\end{proof}

It is worth noting that the left-hand side of Eq.~(\ref{eq:proof1}) is related to the ``sandwiched'' relative Renyi entropy
\beq
\tilde{S}_{\alpha}(\tau\|\rho):=\frac{1}{\alpha-1}\ln{\rm Tr}\left(\rho^{-\frac{\alpha-1}{2\alpha}}\tau\rho^{-\frac{\alpha-1}{2\alpha}}\right)^{\alpha}
\eeq
introduced in the context of quantum information~\cite{Muller2013,Wilde2014}.
By using this quantity, we also have
\beq
I_{\tau}(x)\geq\frac{\alpha-1}{\alpha}\left[I_{\rho}(x)-\tilde{s}_{\alpha}(\tau\|\rho)\right],
\eeq
where $\tilde{s}_{\alpha}$ is defined similarly to $s_{\alpha}$.
Since $\tilde{s}_{\alpha}\leq s_{\alpha}$, this inequality is tighter than the inequality~(\ref{eq:main}).

\sectionprl{Relation to the previous works}
As for the equivalence of the microcanonical and the canonical ensembles,
the vanishing specific relative entropy (the measure equivalence) was extensively studied in Ref.~\cite{Lewis1994} for classical lattice systems.
The relation among the three levels of the equivalence of ensembles was completely clarified by Touchette~\cite{Touchette2015}.

Our understanding of the equivalence of general states $\rho$ and $\tau$ is still not satisfactory.
The result towards this direction was recently obtained in Ref.~\cite{Brandao-Cramer_arXiv2015}, where it was shown that for $d$-dimensional quantum spin systems, if $\rho_N$ has the property of the exponential decay of correlations, $S(\tau_N\|\rho_N)=\mathcal{O}(N^{d/(d+1)})$ implies ${\rm Tr}\rho_N X\approx{\rm Tr}\tau_N X$ for any local intensive operator $X=(1/N)\sum_{i=1}^NO_i$.

The assumption of the exponential decay of correlations excludes the states with strong long-range correlations such as equilibrium states of long-range interacting systems~\cite{Campa_review2009} and some nonequilibrium steady states~\cite{Derrida2002,Cohen-Mukamel2012}.
However, in deriving our result~(\ref{eq:main}), the assumption of exponential decay of correlations is not necessary.
Moreover, the vanishing specific relative entropy $S(\tau_N\|\rho_N)=o(N)$ is weaker than the criterion of $S(\tau\|\rho)=\mathcal{O}(N^{d/(d+1)})$.

\sectionprl{Large-deviation probability of the microcanonical and the canonical ensembles}
We have shown that the measure equivalence implies the macrostate equivalence (at least partial equivalence) from the inequality~(\ref{eq:main}) for $\alpha\rightarrow 1$.
The inequality~(\ref{eq:main}) for $\alpha>1$ also contains useful information on the large-deviation rate function.
We apply the inequality~(\ref{eq:main}) with $\alpha=+\infty$ for $\tau=\rho_{\rm mc}$ and $\rho=\rho_{\rm can}$, where $\rho_{\rm mc}$ is the microcanonical density matrix, $\rho_{\rm mc}=\mathsf{P}_{H\in[E,E+\Delta E)}e^{-S(E)}$, where $S(E)=\ln{\rm Tr}\mathsf{P}_{H\in[E,E+\Delta E)}$ is the microcanonical entropy, and $\rho_{\rm can}$ is the canonical density matrix, $\rho_{\rm can}=e^{\beta (F(\beta)-H)}$, where $e^{-\beta F(\beta)}={\rm Tr}e^{-\beta H}\equiv Z$ is the partition function with $F(\beta)$ the free energy. The inverse temperature is denoted by $\beta$.
We can calculate $S_{\infty}(\rho_{\rm mc}\|\rho_{\rm can})$ as
$S_{\infty}(\rho_{\rm mc}\|\rho_{\rm can})=\ln\max_{i:E_i\in[E,E+\Delta E)}[e^{S(E)-\beta F(\beta)+\beta E_i}]\leq S(E)-\beta F(\beta)+\beta E+\beta\Delta E$.
When the microcanonical and the canonical ensembles are thermodynamically equivalent, $S(E)-\beta F(\beta)+\beta E=o(N)$ and we can choose $\Delta E=o(N)$.
Therefore, $s_{\infty}(\rho_{\rm mc}\|\rho_{\rm can})=0$.
From the inequality~(\ref{eq:main}), we obtain
\beq
I_{\rho_{\rm mc}}(x)\geq I_{\rho_{\rm can}}(x).
\label{eq:large_dev}
\eeq
It is known in classical statistical mechanics that the fluctuation in the microcanonical ensemble is always not greater than that in the canonical ensemble~\cite{Lebowitz1967}.
Equation~(\ref{eq:large_dev}) indicates that not only the small fluctuation, but also the large deviation probability in the microcanonical ensemble is not greater than that in the canonical ensemble~\footnote
{As an exception, when $I_{\rho_{\rm mc}}(x)=I_{\rho_{\rm can}}(x)$, the probability of having $x$ in the microcanonical ensemble can be larger than that in the canonical ensemble.}.
This result is rather trivial, and Eq.~(\ref{eq:large_dev}) can be derived more directly without using the relative entropy, see, for example, Ref.~\cite{Tasaki2016_typicality}.
The purpose here is to demonstrate that Eq.~(\ref{eq:main}) for large values of $\alpha$ contains much information on the rate functions.

\sectionprl{Thermalization}
We consider a pure initial state $|\psi(0)\>=\sum_nc_n|\phi_n\>$, where $|\phi_n\>$ is an eigenstate of the Hamiltonian $H$ with energy $E_n\in[E,E+\Delta E)$.
We assume that there is no degeneracy in $H$.
The system evolves with time under the Hamiltonian $H$, and the state at time $t$ is given by $|\psi(t)\>=\sum_nc_ne^{-iE_nt}|\phi_n\>$.
The stationary state after a sufficiently long time is described by the diagonal ensemble $\rho_D$~\cite{Rigol2008}, which is the long-time average of $|\psi(t)\>\<\psi(t)|$:
\beq
\rho_D=\sum_n|c_n|^2|\phi_n\>\<\phi_n|.
\eeq
We compare it with the microcanonical ensemble $\rho_{\rm mc}$.
For simplicity, we assume that the typical value of $X$ in $\rho_{\rm mc}$ is unique: $\mathcal{E}_{\rho_{\rm mc}}=\{x_{\rm eq}\}$, where $x_{\rm eq}=\<X\>_{\rho_{\rm mc}}$.
We introduce the precision $\delta$ of measuring a macrovariable $X$.
We shall say that the system thermalizes in the precision $\delta$ if $|\<X\>_{\rho_D}-x_{\rm eq}|<\delta$.
For a given $\delta$, there exists $\gamma>0$ defined as
$\gamma:=\inf_{x:|x-x_{\rm eq}|>\delta}I_{\rho}(x)$.
Then, Eq.~(\ref{eq:macro_alpha}) implies that if 
\beq
s_{\alpha}(\rho_D\|\rho_{\rm mc})<\gamma,
\label{eq:equilibrium}
\eeq
the system thermalizes in the precision $\delta$.
Moreover, the inequality~(\ref{eq:main}) ensures that the probability of obtaining an out-of-equilibrium value of $X$ is exponentially small with respect to the system size.
From this observation, by following the argument by Tasaki~\cite{Tasaki2016_typicality}, one can show that, not only the time average, but also an instantaneous value of $X$ is almost certainly in equilibrium with precision $\delta$ for almost all times $t$.
Therefore, the long-time average is not necessary, and the system thermalizes in the sense that 
\beq
\<\psi(t)|\mathsf{P}_{|X-x_{\rm eq}|>\delta}|\psi(t)\><e^{-\nu N}
\eeq
for some $\nu>0$ and sufficiently large $N$ for almost all times $t$, when the condition~(\ref{eq:equilibrium}) is fulfilled.

For $\alpha=2$, the condition~(\ref{eq:equilibrium}) yields
\beq
D_{\rm eff}\geq De^{-\eta N} \text{ with some } 0<\eta<\gamma,
\label{eq:condition2}
\eeq
where $D$ is the dimension of the Hilbert subspace spanned by the energy eigenstates within the microcanonical energy shell,
and $D_{\rm eff}:=(\sum_n|c_n|^4)^{-1}$ is called the inverse participation ratio, which characterizes how many energy eigenstates join to constitute the initial state.
The condition~(\ref{eq:condition2}) was obtained in Ref.~\cite{Tasaki2016_typicality}.

We can obtain a milder sufficient condition for themalization by considering $\alpha\rightarrow 1$.
The condition~(\ref{eq:equilibrium}) then implies
\beq
e^{S(\rho_D)}\geq De^{-\eta N} \text{ with some } 0<\eta<\gamma,
\label{eq:condition1}
\eeq
where $S(\rho)=-{\rm Tr}\rho\ln\rho$ is the von Neumann entropy.
Because $e^{S(\rho)}\geq D_{\rm eff}$, Eq.~(\ref{eq:condition1}) is a milder condition on the initial state than Eq.~(\ref{eq:condition2}).

\sectionprl{Timescale of quantum dynamics}
We prepare the canonical ensemble of the initial Hamiltonian $H_0$, $\rho(0)=e^{-\beta H_0}/{\rm Tr}e^{-\beta H_0}$.
After the system is disconnected from the thermal bath, we change the Hamiltonian $H(t)$, which depends on time in general for $t>0$.
The quantum state at time $t$ is given by $\rho(t)=U_t\rho(0)U_t^{\dagger}$, where $U_t=\mathcal{T}e^{-i\int_0^tH(s)ds}$ is the time-evolution unitary operator with $\mathcal{T}$ the time-ordering operator.

By using the fact that the von Neumann entropy is invariant under the unitary time evolution, the relative entropy between $\rho(t)$ and $\rho(0)$ is obtained as
\beq
S(\rho(t)\|\rho(0))=\beta\left(\<H_0\>_{\rho(t)}-\<H_0\>_{\rho(0)}\right).
\label{eq:timescale}
\eeq
As long as $S(\rho(t)\|\rho(0))/N$ is very small, $\rho(t)$ and $\rho(0)$ predict almost identical expectation values of macrovariables.
Therefore, Eq.~(\ref{eq:timescale}) gives an estimate of timescale of the dynamics of macrovariables.

As an example, let us consider a periodically driven spin system $H(t)=H_0+V(t)$ with $V(t)=V(t+T)$, where $T=2\pi/\omega$ is the period of driving field ($\omega$ is the frequency).
When the driving field is sufficiently fast, $\omega\gg g$, where $g$ is the largest energy scale of single spin, it is rigorously proven that the energy absorption is exponentially slow, $|\<H_0\>_{\rho(t)}-\<H_0\>_{\rho(0)}|/N\lesssim e^{-\mathcal{O}(\omega/g)}t$~\cite{Mori2016_rigorous, Kuwahara2016_Floquet, Abanin_arXiv2015_asymptotic, Abanin_arXiv2015_effective}.
Equation~(\ref{eq:timescale}) implies that the specific relative entropy remains very small and hence, not only the energy, but also any other macrovariables cannot largely change up to an exponentially long timescale, $t\lesssim e^{\mathcal{O}(\omega/g)}$.
Thus, the state $\rho(0)=e^{-\beta H_0}/{\rm Tr}e^{-\beta H_0}$ is regarded as a quasi-stationary state for high frequencies~\cite{Mori2016_rigorous}.

This estimate of timescale should be contrasted with that by the trace distance between $\rho(t)$ and $\rho(0)$, $D(t)={\rm Tr}|\rho(t)-\rho(0)|$.
When $D(t)\ll 1$, it is ensured that two states $\rho(t)$ and $\rho(0)$ are almost indistinguishable.
However, this quantity increases very rapidly especially for large $N$.
This does not necessarily mean that the timescale becomes fast as $N$ increases because there is no distinction between microvariables and macrovariables here.
When we are interested in the time evolution of macrovariables, $S(\rho(t)\|\rho(0))/N$ is a proper choice as a measure of distance between $\rho(t)$ and $\rho(0)$.

\sectionprl{Second law with strict irreversibility in a quantum quench}
As the most nontrivial application of the result, we shall discuss the second law in a quantum quench in an isolated quantum system.
The initial Hamiltonian is denoted by $H_i$, and at time $t=0$, the Hamiltonian is suddenly quenched to $H_f$.
We assume that the initial state is an equilibrium state, $\rho_i^{\mathrm{eq}}=e^{-\beta H_i}/\mathrm{Tr}e^{-\beta H_i}$.
After a quench, the density matrix at time $t>0$ is given by $\rho(t)=e^{-iH_ft}\rho_i^{\mathrm{eq}}e^{iH_ft}$.
We consider the thermodynamic entropy~\cite{Tasaki2016} defined as
\beq
S_{\mathrm{TD}}(\rho)=S(\rho^{\mathrm{eq}}),
\eeq
where $S(\cdot)$ again denotes the von Neumann entropy and $\rho^{\mathrm{eq}}=e^{-\beta(\rho)H}/\mathrm{Tr}e^{-\beta(\rho)H}$ with $\beta(\rho)$ determined by $\<H\>_{\rho}=\<H\>_{\rho^{\mathrm{eq}}}$ is the equilibrium density matrix corresponding to $\rho$~\footnote
{The thermodynamic entropy $S_{\mathrm{TD}}(\rho)$ of the state $\rho$ is also expressed as \unexpanded{$S_{\mathrm{TD}}(\rho)=\sup_{\beta>0}\left[\beta\< H\>_{\rho}+\ln Z(\beta)\right]$}, where $Z(\beta)=\mathrm{Tr}e^{-\beta H}$.}.
The equilibrium density matrix corresponding to the state after the quench is denoted by $\rho_f^{\mathrm{eq}}=e^{-\beta_fH_f}/\mathrm{Tr}e^{-\beta_fH_f}$, where $\beta_f$ is determined so that $\<H_f\>_{\rho(t)}=\<H_f\>_{\rho_f^{\mathrm{eq}}}$ ($\beta_f$ is time-independent because $\<H_f\>_{\rho(t)}$ is time-independent).
Although the von Neumann entropy is invariant under a unitary time evolution, the thermodynamic entropy can vary in an adiabatic process. For several definitions of entropy, see Supplemental Material of Ref.~\cite{Tasaki2016}.

The relative entropy $S(\rho_i^{\mathrm{eq}}\|\rho_f^{\mathrm{eq}})$ is calculated as
\begin{align}
S(\rho_i^{\mathrm{eq}}\|\rho_f^{\mathrm{eq}})&=S(\rho_f^{\mathrm{eq}})-S(\rho_i^{\mathrm{eq}})
\nonumber \\
&\equiv S_{\mathrm{TD}}^{(f)}-S_{\mathrm{TD}}^{(i)},
\label{eq:relative_eq}
\end{align}
and thus the right hand side of Eq.~(\ref{eq:relative_eq}) is nothing but the change of the thermodynamic entropy in the quantum quench.

From the non-negativity of the relative entropy, we obtain $S_{\mathrm{TD}}^{(f)}-S_{\mathrm{TD}}^{(i)}\geq 0$, i.e., the second law of thermodynamics in a quantum quench, which is well known~\cite{Tasaki_arXiv2000}.

However, this result only tells us that the thermodynamic entropy is non-decreasing, and an unsolved crucial problem in the microscopic derivation of the irreversibility is to show that the entropy considerably increases for a sudden quench.
We can give an answer to this problem.
If the initial equilibrium state $\rho_i^{\mathrm{eq}}$ and the final equilibrium state $\rho_f^{\mathrm{eq}}$ are macroscopically different, the relative entropy $S(\rho_i^{\mathrm{eq}}\|\rho_f^{\mathrm{eq}})$ \textit{must be} extensively large~\footnote
{More specifically, by putting $\tau=\rho_i^{\mathrm{eq}}$ and $\rho=\rho_f^{\mathrm{eq}}$, and assuming \unexpanded{$\mathcal{E}_{\tau}=\{x_i\}$} and \unexpanded{$\mathcal{E}_{\rho}=\{x_f\}$} with $x_i\neq x_f$, we obtain \unexpanded{$s(\tau\|\rho)\geq I_{\rho}(x_i)>0$} from Eq.~(\ref{eq:macro}), which implies \unexpanded{$S(\tau\|\rho)\geq NI_{\rho}(x_i)=\mathcal{O}(N)$}.}.
Hence, for a nontrivial quench, in which the initial equilibrium state and the final equilibrium state are distinct, we obtain
\beq
S_{\mathrm{TD}}^{(f)}-S_{\mathrm{TD}}^{(i)}=\mathcal{O}(N),
\label{eq:second_law}
\eeq
which shows the thermodynamic entropy increases extensively.
If the increase of the entropy is not extensive, it implies that $\rho_i^{\mathrm{eq}}$ and $\rho_f^{\mathrm{eq}}$ are macrostate equivalent, and thus such a quantum quench is trivial.

Equation~(\ref{eq:second_law}) shows that the thermodynamic entropy exhibits the \textit{strict irreversibility} in a quantum quench, and an extensive increase of the thermodynamic entropy is a general result solely derived from the ensemble \textit{non}-equivalence of the initial and the final states.
Thus, the result of this paper would provide us a more satisfactory microscopic explanation on the second law and the irreversibility of a macroscopic system.

\sectionprl{Conclusion}
In this work, we have made clear the relation between the vanishing specific relative entropy in the thermodynamic limit (the measure equivalence) and the macrostate equivalence for general states $\rho$ and $\tau$ by deriving the inequality connecting the specific relative entropy to the large-deviation rate function of a macrovariable.
When the typical value of a macrovariable is unique, this inequality shows that the two ensembles are macrostate equivalent whenever the specific relative entropy vanishes in the thermodynamic limit.
When the typical value is not unique, there are two possibilities, partial equivalence or full equivalence. 
Moreover, the inequality~(\ref{eq:main}) provides us useful information on the rate functions of macrovariables.
As applications, we have discussed the problem of thermalization, the timescale of quantum dynamics, and the second law in a quantum quench.
Since the inequality~(\ref{eq:main}) holds both for quantum and classical systems, it has a potential usefulness for, e.g., an efficient construction of the generalized Gibbs ensemble in integrable systems~\cite{Sels-Wouters2015}, the ensemble equivalence in networks~\cite{Squartini2015}, and hopefully, some nonequilibrium problems such as statistical ensembles for nonequilibrium systems~\cite{Evans2004,Evans2005} and path ensembles of nonequilibrium Markov processes~\cite{Chetrite-Touchette2013,Chetrite-Touchette2014}.
Especially, the inequality~(\ref{eq:main}) contains much more information than the non-negativity of the relative entropy, which allows us to derive strict irreversibility in a quantum quench.
Therefore, by utilizing the result obtained in this work, one will considerably strengthen several results obtained by the non-negativity of the relative entropy~\cite{Sagawa_lecture2013}, and exploring such applications is a fascinating future problem.

\begin{acknowledgments}
I wish to thank Hugo Touchette and Mark M. Wilde for useful comments.
This work was financially supported by JSPS KAKENHI Grant No.~15K17718.
\end{acknowledgments}

\end{document}